\documentclass[%
 reprint,
 superscriptaddress,
 amsmath,amssymb,
 aps,
 prl,
 longbibliography,
lengthcheck,%
]{revtex4-1}

\usepackage[dvipdfmx]{graphicx}
\usepackage{dcolumn}
\usepackage{bm}
\usepackage{hyperref}
\usepackage{color}
\usepackage{ulem}

\begin{document}

\preprint{APS/123-QED}

\title{
  Singular nonlinear response in metals due to divergent Berry curvature:\\
  application to magneto-resistance in metals with type-II Weyl nodes
}

\author{Hiroaki Ishizuka}
\affiliation{
Department of Applied Physics, The University of Tokyo, Bunkyo, Tokyo, 113-8656, JAPAN 
}

\author{Naoto Nagaosa}
\affiliation{
Department of Applied Physics, The University of Tokyo, Bunkyo, Tokyo, 113-8656, JAPAN 
}
\affiliation{
RIKEN Center for Emergent Matter Science (CEMS), Wako, Saitama, 351-0198, JAPAN
}

\date{\today}

\begin{abstract}
The physical properties of metals are often given by the sum of the contributions from the electrons consisting the Fermi surface (FS), and therefore, fine structures of the electronic bands and Bloch functions are often masked by the integral over FS. As a consequence, usually, the singular structures in the electronic bands are often not reflected to the macroscopic quantities. In this work, we investigate the anomaly-related magnetoresistance in metals with type-II Weyl nodes close to the FS, and find that the anomaly-related current increases divergently, showing a singular structure. Detailed analysis on a simple model with multiple Weyl nodes shows that the contribution to the magnetoresistance is dominated by the electrons in the vicinity of the Weyl nodes; this is related to the fact that the current is given by the square of the Berry curvature, which enhances the contribution from the electrons around the Weyl nodes. The above results potentially allows an estimate of the anomaly-induced current without precise information of the entire Band structure.
\end{abstract}

\pacs{
}

\maketitle

{\it Introduction} --- In the quantum theory of solids, the physical properties of metals are often given by the sum of contributions from the electrons consisting the Fermi surface (FS) or the electrons inside it. For instance, electronic specific heat and longitudinal conductivity are related to the density of states at the FS. This aspect of the physical properties of the metals often masks the fine structure of the electronic bands, and therefore, the fine structures are difficult to be observed.

The properties related to Berry curvature $\bm b_{\bm p}$ is also often given by the integral over the Fermi volume or FS. For example, the Hall conductivity is related to the sum of Berry curvature of the Fermi volume~\cite{Thouless1982,Kohmoto1985}. Nevertheless, it was pointed out that the anomalous Hall effect is sensitive to the band crossings where the $\bm b_{\bm p}$ is enhanced singularly; the temperature and Fermi energy dependence of the Hall conductivity in several oxides is interpreted as a consequence of the distribution of Berry curvature~\cite{Fang2003,Takahashi2018}. However, it is given by the integral of $\bm b_{\bm p}$ over the Fermi volume or Fermi surface, which often weaken the singular nature of $\bm b_{\bm p}$.

On the other hand, recently, it was pointed out that the Berry curvature also contributes to nonlinear responses, such as photocurrent~\cite{Moore2010,Sodemann2015,Ishizuka2016}, second-harmonic generation~\cite{Sodemann2015,Ishizuka2017}, and in linear~\cite{Sharma2017,Sekine2018} and quadratic~\cite{Son2013} magnetoresistance (MR); these phenomena are potentially relevant to transport phenomena in Weyl semimetals (WSM)~\cite{Herring1937,Murakami2007,Burkov2011,Xu2011,Wan2011,Lv2015-2,Xu2015}, which are studied extensively in material-based calculations~\cite{Zhang2018a,Zhang2018b} and in experiments~\cite{Liang2014,Huang2015,Xiong2015,Li2016,Zhang2016,Wu2016,Ma2017,Osterhoudt2017}. The theoretical formula given in many of the theoretical proposals are also related to some form of the integrals of $\bm b_{\bm k}$. Therefore, the above argument is expected to hold for these phenomena. On the other hand, the leading order in the quadratic MR is proportional to the sum of the square of the Berry curvature~\cite{Son2013,Cortijo2018,Ishizuka2018}. Therefore, the MR may reflect the singular structure of the Berry curvature even when the Weyl nodes are part of a large Fermi surface, i.e., in type-II WSM~\cite{Soluyanov2015}. 

In this work, we discuss that response coefficients related to the Berry curvature on FS generally induce a singular structure when the response is related to square or a higher power of $\bm b_{\bm p}$. The argument is based on a semiclassical Boltzmann theory~\cite{Sundaram1999,Xiao2010}. As an example, we revisit the Weyl Hamiltonian with tilting and a metal with two type-II Weyl nodes, and study the longitudinal MR which was recently studied by different methods~\cite{Soluyanov2015,Sharma2017,Wei2018}. We find that, in the semiclassical limit, the magnetic-field induced current increases as the chemical potential $\mu$ approaches the Weyl node even for the metals; the current is proportional to $\mu^{-2}$ near the nodes. We also find that the magnetic-field-induced current in type-II WSM can be enhanced by the tilting, possibly become more than an order of magnitude larger than the type-I WSM with a similar velocity.   

{\it Semiclassical theory} --- A semiclassical approach to study the chiral anomaly and related transport phenomena was recently proposed~\cite{Son2013}, and was extended to more general cases, including weak-localization~\cite{Kim2014} and extention to general Hamiltonian~\cite{Lundgren2014,Sharma2016}. In this work, however, we take a slightly different approach using a recently introduced formula for the response in the order of ${\cal O}(EB^2)$~\cite{Ishizuka2018}:
\begin{align}
  \bm J_b^{(2)}=& -e^4\tau\int \frac{dp^3}{(2\pi)^3}\left[\bm W_{\bm p}(\bm E\cdot\bm W_{\bm p})\right](f_{\bm p}^0)',\label{eq:Jb2}
\end{align}
where $e<0$ is the electron charge, $\tau$ is the relaxation time, $\bm v_p$ is the velocit of electrons with momentum $\bm p$, and $\bm W_{\bm p}\equiv\bm b_{\bm p}\times(\bm v_{\bm p}\times\bm B)$. The formula in Eq.~\ref{eq:Jb2} corresponds to expanding the phase-space volume factor $D\equiv(1-e\bm B\cdot\bm b_{\bm p})^{-1}$ in the frequently used formula~\cite{Kim2014,Lundgren2014,Sharma2016} with respect to $\bm B$.

\begin{figure}
  \includegraphics[width=\linewidth]{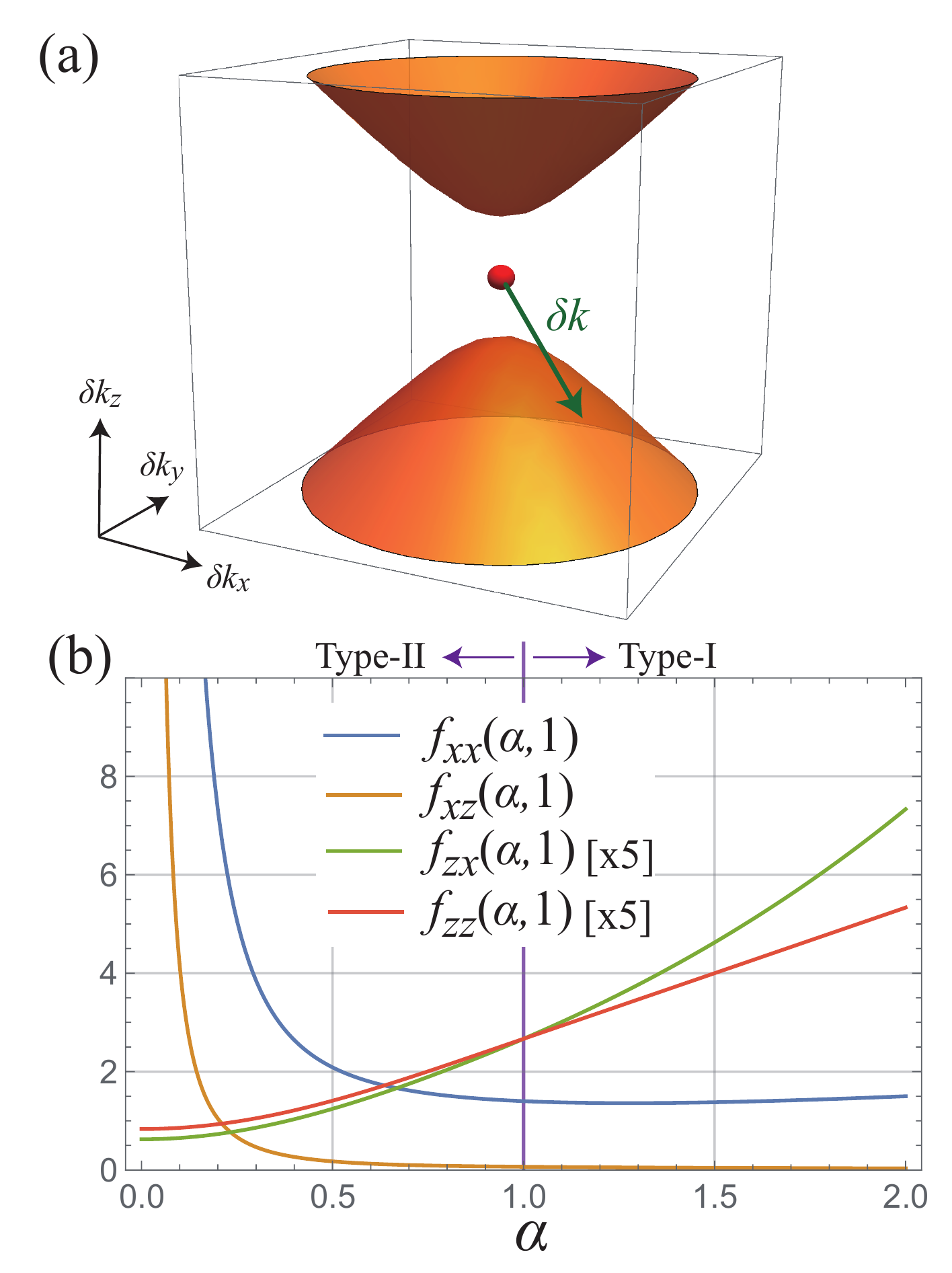}
  \caption{(Color online) Electronic structure of type-II Weyl fermion. (a) Fermi surface around type-II Weyl node (shown in shaded surfaces). The sphere at the center is the Weyl node and the arrow indicates $\bm {\delta k}$. (b) Plot of $f_{ab}(\alpha,\beta=1)$ with respect to $\alpha=v_z/v_0$; the magnetoconductivity reads $J_a\propto v_0^3f_{ab}(v_z/v_0,v_\perp/v_0)E_aB_b^2/\mu^2$ ($a,b=x,z$); here, $z$ is the direction of the tilting.}
  \label{fig:weyl_node}
\end{figure}

Equation~\eqref{eq:Jb2} shows that the Berry-phase contribution (anomaly-related contribution) to the longitudinal magnetoresistance (MR) is always negative MR. Suppose we apply the electric field along $\hat{\bm e}$ ($\bm E=E\hat{\bm e}$) and measure the current along the same direction. Then the current reads
\begin{align}
  \hat{\bm e}\cdot\bm J_b^{(2)}=& -e^4\tau E\int \frac{dp^3}{(2\pi)^3}\left(\hat{\bm e}\cdot\bm W_{\bm p}\right)^2(f_{\bm p}^0)'>0,\label{eq:Jb2-2}
\end{align}
indicating that the current is always zero or positive along the direction of $\bm E$. Another interesting aspect is that the contribution from the electrons with momentum $\bm p$ is proportional to $W_{\bm p}^2\sim b_{\bm p}^2$. For the electrons close to a Weyl node, this implies $W_{\bm p}^2\sim \delta k^{-4}$ where $\bm {\delta k}\equiv\bm k-\bm k_W$ and $\bm k_W$ is the location of the Weyl node [See Fig.~\ref{fig:weyl_node}(a)]. Therefore, the contribution to the MR decays rapidly with increasing $|\bm {\delta k}|$.

To be more quantitative, we consider a generalized type-II Weyl Hamiltonian $H=R_0(\bm p)+\sum_{a=x,y,z}R_a(\bm p)\sigma^a$ where $R_0$ and $R_a$ are a power series of $p_a$ and $\sum_{a=x,y,z}R_a^2(\bm p)=0$ only at $\bm p=\bm 0$; in the below, we call the bands with eigenenergy $\varepsilon_{\bm p\pm}=R_0\pm\sqrt{R_x^2+R_y^2+R_z^2}$ as $\pm$ bands. We further assume that, using the polar coordinate, the Fermi surface of this model is given by $(p,\theta_\pm(p,\phi),\phi)$ where $\theta_\pm(p,\phi)$ is a single-valued function that determines the Fermi surface of the + and - bands and $\theta_+(p,\phi)>\theta_-(p,\phi)$. This essentially assumes the tilting is along $z$ axis and the energy monotonically increases about $p_z$, and the two bands has one Fermi surface which extends to $p\to\infty$. Then, an integral of a function $F(\bm p)$ over the Fermi surface reads
\begin{align}
\int \frac{dp^3}{(2\pi)^3}&F_\pm(\bm p) \delta(\varepsilon_{\bm p\pm}-\mu)\nonumber\\
  &\propto\int dpd\phi\left.\frac{pF_\pm(\bm p)}{|\bm n_\theta\cdot\bm v_{\bm p\pm}|}\right|_{\theta=\theta_\pm(p,\phi)},\label{eq:Fint2}
\end{align}
where $\bm n_\theta$ is a unit vector along the $\theta$ axis. Assuming $\varepsilon_{\bm p\pm}\propto p^{\tilde\eta}$ and $F_{\pm}(\bm p)\propto p^{-\tilde a}$ at $p\to\infty$, the integrand become $\propto p^{2-\tilde a-\tilde\eta}\left.g_\pm(\theta,\phi)\right|_{\theta=\theta_\pm(p,\phi)}$, where $g$ is a function of $\theta$ and $\phi$. Hence, the $p\gg1$ part of the integral in Eq.~\eqref{eq:Fint2} converges when $\tilde a>3-\tilde\eta$; this implies that the contribution from $p\gg1$ part only have a limited contribution. Similarly, if $\varepsilon_{\bm p\pm}\propto p^{\eta}$ and $F_{\pm}(\bm p)\propto p^{-a}$ when $p\ll1$, the $p\ll1$ part of the integral diverges when $a>3-\eta$ if the Fermi level is at the node; the integral remain finite when the Fermi level is away from the node and diverges as it approach the node. In case of the type-II Weyl node, $\varepsilon_{\bm p\pm}\propto p$ and $\bm b_{\bm p\pm}\propto p^{-2}$ for both $p\ll1$ and $p\gg1$. Therefore, for Eq.~\eqref{eq:Jb2}, $F(\bm k)\propto k^{-4}$ and satisfies $4=a>3-\eta=2$. On the other hand, if a response is linearly proportional to $\bm b_{\bm p\pm}$, then $a=2$ and it implies the contribution away from the nodes are also important. This argument is consistent with the previous theoretical results on the MR which discovered that the behavior of MR connects smoothly across the phase transition between type-I and type-II WSM~\cite{Sharma2017}.

{\it Tilted Hamiltonian} --- We first consider a type-II Weyl Hamiltonian
\begin{align}
H_{W2}=v_\perp k_x\sigma^x+v_\perp k_y\sigma^y+v_zk_z\sigma^z+v_0 k_z\sigma^0,\label{eq:HW2}
\end{align}
where $\sigma^a$ ($a=x,y,z$) is the Pauli matrices and $\sigma^0\equiv\text{diag}(1,1)$ is the $2\times2$ unit matrix. By applying Eq.~\ref{eq:Jb2}, the current induced by the longitudinal magnetic field along $a$ axis reads
\begin{align}
J_a=\frac{\sigma_0 v_0^3}{\mu^2}f_{ab}(v_z/v_0,v_\perp/v_0)E_aB_b^2
\end{align}
with $a,b=x,y,z$, where $\sigma_0=q^4\tau/(8\pi^2)$ is the coefficient for the type-I Weyl node with velocity $v=1$~\cite{Son2013} and
\begin{subequations}
\begin{align}
f_{xx}(\alpha,\beta)=&\beta^2\frac{3\alpha^8-7\alpha^6+25\alpha^4+255\alpha^2+60}{240\alpha^2},\label{eq:fxalpha}\\
f_{xz}(\alpha,\beta)=&\beta^4\frac{-2\alpha^6+5\alpha^4+5}{120\alpha^2},\\
f_{zx}(\alpha,\beta)=&\frac{-2\alpha^8+11\alpha^6-25\alpha^4+65\alpha^2+15}{120\alpha^2},\\
f_{zz}(\alpha,\beta)=&\beta^2\frac{\alpha^6-5\alpha^4+15\alpha^2+5}{30},\label{eq:fzalpha}
\end{align}
\end{subequations}
when $\alpha<1$ and
\begin{subequations}
\begin{align}
f_{xx}(\alpha,\beta)=&\beta^2\frac{8\alpha^1+13}{15\alpha},\\
f_{xz}(\alpha,\beta)=&\beta^4\frac{1}{15\alpha},\\
f_{zx}(\alpha,\beta)=&\frac{\alpha^3+7\alpha}{15},\\
f_{zz}(\alpha,\beta)=&\beta^2\frac{\alpha}{15},
\end{align}
\end{subequations}
when $\alpha>1$. The results for $y$ is the same as $x$, due to the rotational symmetry about $z$ axis. Similar to the type-I WSM, the current proportional to $EB^2$ increase with $\mu^{-2}$ as it approaches the Weyl node. Although the Weyl node consists only a small part of the Fermi surface in lattice models, this divergent behavior is expected to survive as we discussed in the previous section. We confirm this in the next section. It is interesting that the above result seems to imply the MR is related to the Weyl node even in the semiclassical regime where the chiral anomaly argument in the quantum limit~\cite{Neilsen1983,Fukushima2008} does not directly apply.

Another interesting feature is the dependence of the current to the tilting; figure~\ref{fig:weyl_node}(b) plot Eqs.~\eqref{eq:fxalpha}-\eqref{eq:fzalpha}. While the current along $z$ axis remains to be in a similar order to that of the type-I WSM, the current along $x$ increase divergently with increasing tilting, possibly be more than an order of magnitude larger than that of the type-I.

\begin{figure}
  \includegraphics[width=\linewidth]{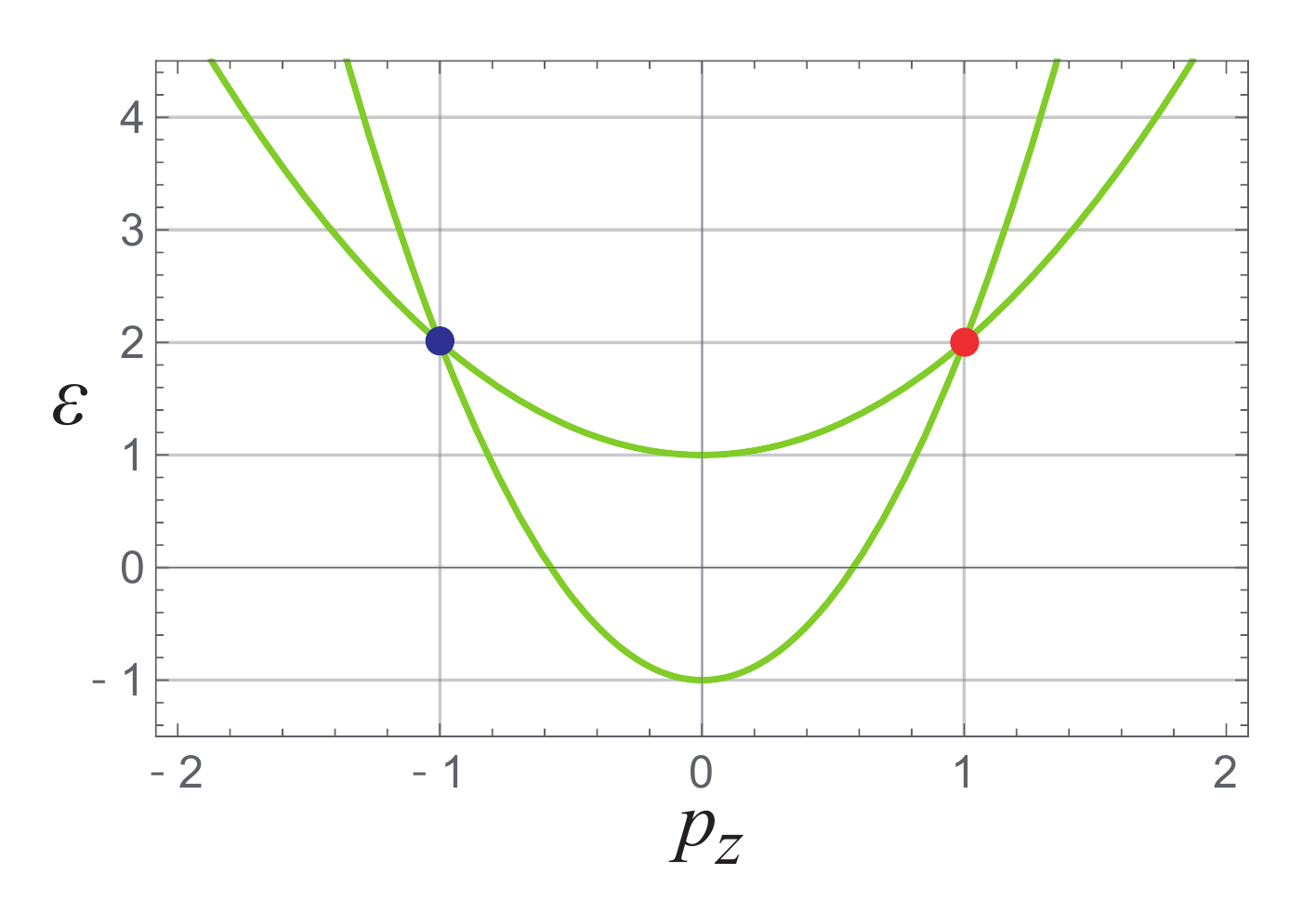}
  \caption{(Color online) Dispersion of the Hamiltonian $H_D$ for $m=1/4$ and $p_0=1$. The two crossings at $p_z=\pm1$ are the Weyl nodes.}
  \label{fig:dipolar}
\end{figure}

{\it Dipolar Model} --- We next test whether the singular response for the type-II Weyl Hamiltonian survives in a model with closed Fermi surfaces, at least when $\mu$ is close to the Weyl nodes. As an example, we here consider a model with two Weyl nodes which we call dipolar model:
\begin{align}
H_D=p_x\sigma^x+p_y\sigma^y+(p_z^2-p_0^2)\sigma^z+\frac{p^2}{2m}\sigma_0,\label{eq:HD}
\end{align} 
where $p^2\equiv p_x^2+p_y^2+p_z^2$. The band structure of this model along $p_x=p_y=0$ line is shown in Fig.~\ref{fig:dipolar}. This model has two Weyl nodes each located at $\bm p=(0,0,\pm p_0)$ with the energy $\mu_W=p_0^2/(2m)$. The nodes are of type-I when $m>1/2$ and become type-II for $|m|<1/2$; in the rest, we focus on the case $0<m<1/2$. The band plotted in Fig.~\ref{fig:dipolar} is for $m=1/4$ and $p_0=1$.

\begin{figure}
  \includegraphics[width=\linewidth]{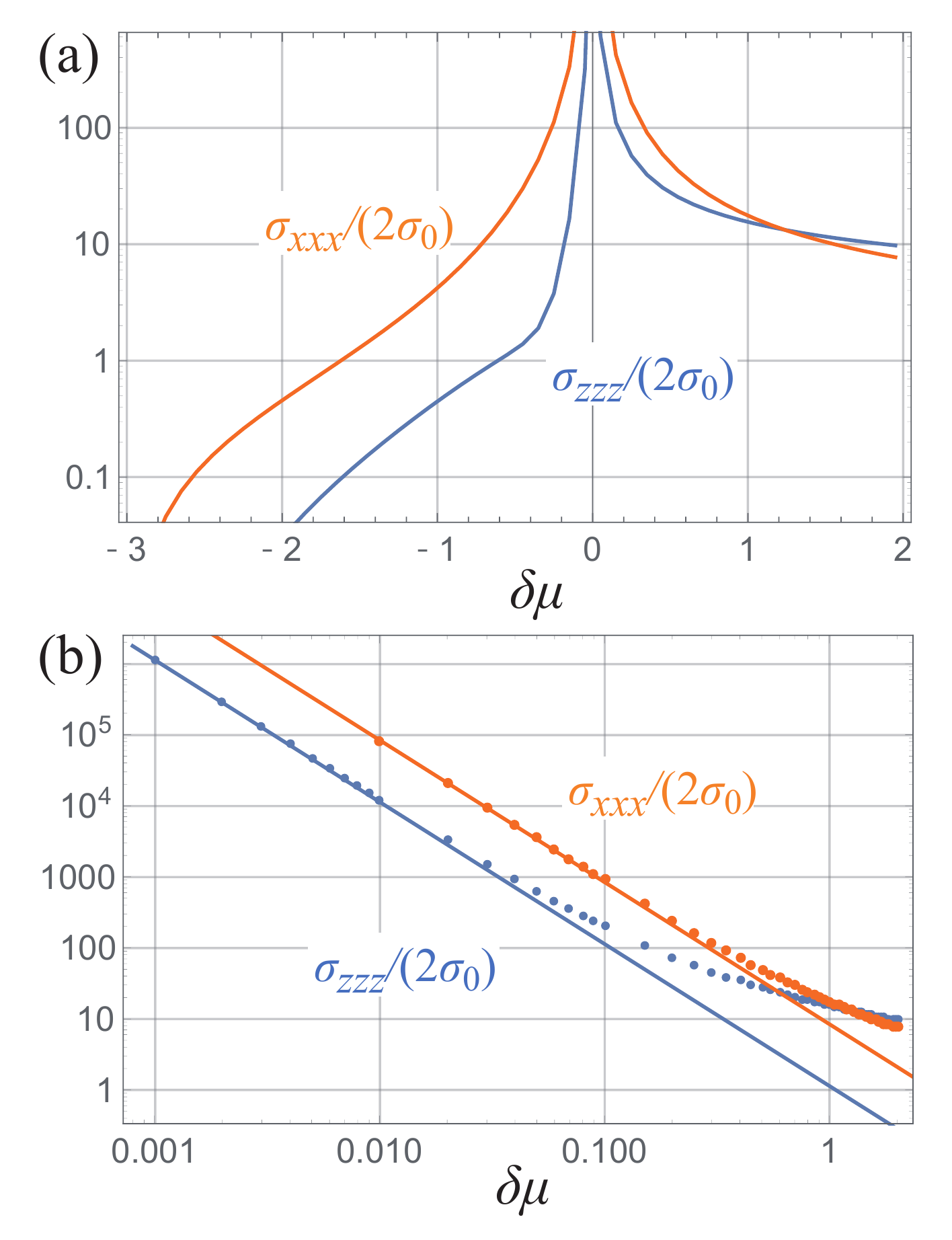}
  \caption{(Color online) Nonlinear conductance for the longitudinal magnetoresistance $(J_b^{(2)})^a=\sigma_{aaa}B_a^2E_a$. (a) Chemical potential $\mu$ dependence of $\sigma_{xxx}/2\sigma_0$ and $\sigma_{zzz}/2\sigma_0$ calculated numerically. (b) The fitting of the numerical results (dots) with $1/(\delta\mu)^2$ ($\delta\mu\equiv\mu-\mu_W$, $\mu_W=2$). The fitted functions are shown by solid lines. All results are for $m=1/4$ and $p_0=1$.}
  \label{fig:saaa}
\end{figure}

To investigate the MR in $H_D$, we numerically calculated the ${\cal O}(EB^2)$ current, i.e., $(J_B^{(2)})^a=\sigma_{bba}B_b^2E_a$, using Eq.~\eqref{eq:Jb2}. The results for $\sigma_{xxx}$ and $\sigma_{zzz}$ are shown in Fig. ~\ref{fig:saaa}(a); in this model, $\sigma_{yyy}$ become the same as $\sigma_{xxx}$ due to the rotational symmetry about $z$ axis. Both results for $\sigma_{xxx}$ and $\sigma_{zzz}$ shows a divergence at $\mu_W=2$ where the chemical potential crosses the Weyl nodes, and the conductivity decays as $\mu$ moves away from the nodes. The conductivity for $x$ is about an order of magnitude larger than that of $z$ axis, consistent with the above argument on the type-II Weyl Hamiltonian.

Figure.~\ref{fig:saaa}(b) shows the fitting of $\sigma_{aaa}$ ($a=x,z$) for $\mu>2$ to a function $h(\mu)=2C\sigma_0/(\mu-\mu_W)^2$, where $C$ is a constant. The results for both $a=x$ and $z$ show a good fit close to $\mu_W=2$ with $C=8.422$ and $C=1.136$ for $\sigma_{xxx}$ and $\sigma_{zzz}$, respectively; the fitting were done for data with $0<\delta\mu<0.1$ where $\delta\mu\equiv\mu-\mu_W$. These numbers of $C$ are in good accordance with the analytic results for the Weyl Hamiltonian in Eq.~\eqref{eq:HW2}. By expanding the model in Eq.~\eqref{eq:HD} around the Weyl point, we find the effective Hamiltonian around the Weyl nodes correspond to Eq.~\eqref{eq:HW2} with $v_\perp=1$, $v_z=\pm2k_0$, and $v_0=\pm p_0/m$. Substituting these relations to Eqs.~\eqref{eq:fxalpha} and \eqref{eq:fzalpha}, we obtain $v_0^3f_{xx}(v_z/v_0)\simeq8.348$ and $v_0^3f_{zz}(v_z/v_0)\simeq1.127$, in good agreement with the fitting for the numerical results of model in Eq.~\eqref{eq:HD}. The results implies that when $\mu$ is sufficiently close to the Weyl nodes ($\delta\mu\lesssim0.1$ in the case of Fig.~\ref{fig:saaa}), the contribution to the magnetoresistance is dominated by the contribution from the Weyl nodes. As a consequence, the longitudinal MR shows a singular structure though the system is a metal with only a part of the Fermi surface has the Weyl fermion features. Also, as implied from Eq.~\eqref{eq:fxalpha}, $\sigma_{xxx}$ for the model in Eq.~\eqref{eq:HD} is about an order of magnitude larger than that expected in a type-I WSMs with the same velocity, $\sigma_{aaa}=\sigma_0/\delta\mu^2$~\cite{Son2013}.

{\it Magnetoresistance in Candidate Materials} --- The above arguments also implies an estimate of the longitudinal MR ratio may be possible just from the effective Weyl Hamiltonian at the node. To investigate this possibility, we estimate the MR ratio of WTe$_2$; as both ohmic and the anomaly-related current are linearly proportional to the relaxation time, the MR ratio is independent of $\tau$ in the semiclassical limit. Using the Drude formula for the Ohmic current $\sigma=\tau e^2 n/m^\ast$, the ratio of conductivity reads
\begin{align}
\chi_{ab} = \frac{\sigma_{ab}}{\sigma}=\frac{m^\ast e^2v_0}{8\pi^2n\hbar}f_{ab}(v_\perp/v_0,v_z/v_0)B_b^2.
\end{align}
Here, we explicitly wrote Planck constant $\hbar$, which was assumed $\hbar=1$ in the above sections. The effective Weyl Hamiltonian for WTe$_2$ was recently given in Ref.~\cite{Soluyanov2015} which finds two quartets of Weyl nodes ($W_1$ and $W_2$), each node in the quartet related by the crystal symmetry; to make an order estimate, we use $v_0=2.8$ eV\AA, $v_\perp=0.5$ eV\AA, and $v_z=0.2$ eV$\rm \AA$ for $W_1$ and $v_0=1.4$ eV\AA, $v_\perp=0.5$ eV\AA, and $v_z=0.2$ eV$\rm \AA$ for $W_2$. The carrier density $n\sim 10^{19}$ cm$^{-3}$~\cite{Zhu2015,Pan2017,Gong2017} and effective mass $m^\ast\sim 0.15 m_e$~\cite{Lv2015}, where $m_e$ is the free electron mass is taken from the experiment. Assuming the chemical potential $\mu\sim10$meV away from the Weyl nodes, we find the largest contribution comes from $\chi_{xx}\sim10^{-3}B^2$; this is roughly consistent with recent experiments which finds $\sim0.1$\% MR ratio with the magnetic field of order $B\sim1$T~\cite{Wang2016,Li2017}.

Regarding the singular structure in the $\mu$ dependence, magnetic WSMs~\cite{Takahashi2018,Wan2011,Kuroda2017} might be a potentially useful setup. Unlike the non-centrosymmetric WSMs, controllability of the existance and the position of the Weyl nodes in magnetic WSM is a potential advantage for studying $\mu$ dependence by moving the Weyl nodes instead of controlling $\mu$. As an example, we here focus on the Weyl nodes in EuTiO$_3$~\cite{Takahashi2018}; we here focus on the pair located close to the $\Gamma$ point. Using the model used in Ref.~\cite{Takahashi2018} and $\sigma\sim10^2$S/cm, we find $\chi_{xx}\sim 10^{-5}B^2$; the smaller ratio comes from smaller velocity (Roughly, the conductivity is proportional to the cubic of velocity.). Interestingly, in our calculation, the conductivity along $z$ axis become relatively large under the magnetic field along $x$ axis $\sigma_{xxz}=10^{-4}B^2$. This is an opposite trend to that in the isotropic Weyl nodes; in this case, $\sigma_{xxz}$ is an order of magnitude smaller than $\sigma_{xxx}$~\cite{Ishizuka2018}.

{\it Linear Magneto-conductivity} --- In a recent work, it was pointed out that the tilting of Weyl cone gives rise to a longitudinal MR which is linearly proportional to the magnetic field~\cite{Sharma2017}. Using the same procedure with Eq.~\eqref{eq:Jb2}, we find the semiclassical formula for linear MR reads
\begin{align}
\bm J_B^{(1)}&= q^2\tau\sum_{\alpha=\pm}\int \frac{dp^3}{(2\pi)^3}\bm W_{\bm p\alpha}(\bm E\cdot\bm v_{\bm p\alpha})(f_{\bm p\alpha}^0)'\nonumber\\
&-q^2\tau\sum_{\alpha=\pm}\int \frac{dp^3}{(2\pi)^3}(\bm B\cdot\bm E)(\bm b_{\bm p\alpha}\cdot\bm v_{\bm p\alpha})\bm v_{\bm p\alpha}(f_{\bm p\alpha}^0)'.\label{eq:Jb1}
\end{align}
In general, this term vanish in a system with time-reversal symmetry. This is shown from the fact that in the time-reversal symmetric systems, $\varepsilon_{\bm p\alpha}=-\varepsilon_{\bm p\alpha}$, $\bm b_{\bm p\alpha}=-\bm b_{\bm p\alpha}$, and $\bm v_{\bm p\alpha}=-\bm v_{\bm p\alpha}$. This is a manifestation of Onsager's reciprocal theorem which states $\sigma_{aa}(\bm B)=\sigma_{aa}(-\bm B)$, where $J_a=\sigma_{aa}(\bm B)E_a$; the Weyl Hamiltonian without tilting happens to possess the above property of $\varepsilon_{\bm p\alpha}$, $\bm b_{\bm p\alpha}$, and $\bm v_{\bm p\alpha}$. Therefore, the current in Eq.~\eqref{eq:Jb1} vanish if no tilting exists. Similarly, in a time-reversal symmetric WSM with tilting, the current in Eq.~\eqref{eq:Jb1} cancels between different nodes. Indeed, a recent semiclassical calculation considering time-reversal symmetric WSM finds only MR that is proportional to $B^2$~\cite{Wei2018}. Therefore, the linear MR is a consequene of time-reversal symmetry breaking. Also, as $a=2$ and $\eta=1$, no singular structure is expected from the $p\to0$ limit.

{\it Discussion} --- To summarize, in this work, we investigated the anomaly-related magnetoresistance in metals with type-II Weyl nodes, focusing on the current of ${\cal O}(EB^2)$. Using a semiclassical transport theory, we find that the anomaly-related current shows a singular structure when the chemical potential is close to the Weyl nodes. We further show that the dominant contribution to the magnetoresistance comes from the Weyl nodes; this is related to the fact that the current is given by the integral over the square of the Berry curvature. In addition, the analysis for the type-II Weyl Hamiltonian shows that the tilting enhances the negative magnetoresistance, sometimes by more than an order of magnitude compared to the type-I Weyl nodes. The above results imply that the magnetoresistance in type-II Weyl semimetals, when the Fermi level is close to the node, is directly related to the Weyl nodes despite the large Fermi surface which most of the surface is not related to Weyl physics. Experimentally, this feature may allow estimating the magnitude of the anomaly-related current only from the effective Weyl Hamiltonian.

This work was supported by JSPS KAKENHI Grant Numbers JP16H06717, JP18H03676, JP18H04222, and JP26103006, ImPACT Program of Council for Science, Technology and Innovation (Cabinet office, Government of Japan), and CREST, JST (Grant No. JPMJCR16F1).

\end{document}